\documentclass[
    superscriptaddress,
    twocolumn,
    a4paper,
    floatfix,
    longbibliography
]{revtex4-1}

\usepackage[american]{babel}
\usepackage[utf8x]{inputenc}

\usepackage{grffile} 

\usepackage{microtype}

\usepackage{bibunits} 

\makeatletter
\newcommand*{\newbibstartnumber}[1]{%
  \apptocmd{\thebibliography}{%
    \global\c@NAT@ctr #1\relax
    \addtocounter{NAT@ctr}{-1}%
  }{}{}%
}
\makeatother

\usepackage{amsmath}
\usepackage{amssymb}
\usepackage{bbm}
\usepackage{mathtools}
\usepackage{dsfont}
\usepackage{braket}
\usepackage{mathrsfs}

\usepackage[clock]{ifsym}

\usepackage{graphicx}
\usepackage[svgnames]{xcolor}

\usepackage{siunitx} 

\usepackage{ragged2e}
\usepackage{array}
\usepackage{tabularx}
\usepackage{booktabs}

\definecolor{cset-aps-blue}{RGB}{18,84,168}
\definecolor{cset-aps-limegreen}{RGB}{153,204,51}

\definecolor{cset-aps-blueberry}{RGB}{28,128,158}
\definecolor{cset-aps-turquoise}{RGB}{0,67,88}
\definecolor{cset-aps-limegreen}{RGB}{190,219,67}
\definecolor{cset-aps-darkblue}{RGB}{31,138,112}
\definecolor{cset-aps-yellow}{RGB}{255,225,25}
\definecolor{cset-aps-orange}{RGB}{253,116,0}
\definecolor{cset-aps-red}{RGB}{219,0,43}

\usepackage{tikz}

\usepackage{pgfplots}
\pgfplotsset{%
    every axis legend/.append style={%
        cells={anchor=west},
        at={(0.96,0.04)},
        anchor=south east,
        font=\scriptsize,
        },
    every axis/.append style={%
        yticklabel style={%
            /pgf/number format/fixed zerofill,
            /pgf/number format/precision=2},
        },
    width= \textwidth,
    height=8cm,
    xmajorgrids=true,
    xminorgrids=false,
    minor x tick num=1,
}
\usepgfplotslibrary{external}

\usetikzlibrary{decorations}

\newcommand{\ii}{\operatorname{i}}
\newcommand{\expUp}[1]{\operatorname{e}^{#1}}
\newcommand{\dd}[1]{\mathrm{d} {#1} \;}

\newcommand{\eg}[0]{\textit{e.\,g.}}
\newcommand{\ie}[0]{\textit{i.\,e.}}
\renewcommand{\bra}[1]{\ensuremath{\langle #1|}}
\renewcommand{\ket}[1]{\ensuremath{|#1\rangle}}
\renewcommand {\braket}[2]{\ensuremath{\langle #1|#2\rangle}}

\newcommand{\Sec}[1]{\vspace{0.5cm}\noindent\textbf{#1}\\}

\usepackage{hyperref}
\hypersetup{%
    colorlinks=true,
    linkcolor={cset-aps-red},
    linkbordercolor={cset-aps-red},
    filecolor={cset-aps-orange},
    filebordercolor={cset-aps-orange},
    citecolor={cset-aps-blue},
    citebordercolor={cset-aps-blue},
    urlcolor={cset-aps-blue},
    urlbordercolor={cset-aps-blue},
    menucolor={cset-aps-limegreen},
    menubordercolor={cset-aps-limegreen},
    breaklinks=true,
    pdfborderstyle={/S/U/W 2},
    pdfpagemode=UseOutlines,
    pdfstartpage={1},
}

\newcommand{\orcid}[1]{\href{https://orcid.org/#1}{\includegraphics[width=7pt]{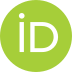}}}

\newcommand{\affTUDa}{\address{Technische Universit{\"a}t Darmstadt, Fachbereich Physik, Institut f{\"u}r Angewandte Physik, Schlossgartenstr. 7, D-64289 Darmstadt, Germany}}

\begin{document}

\title[Title]{A unified theory of tunneling times promoted by Ramsey clocks}

\author{Patrik Schach\,\orcid{0000-0002-6672-9692}}
\affTUDa
    
\author{Enno Giese\,\orcid{0000-0002-1126-6352}}
\affTUDa

\collaboration{This article has been published in \href{https://doi.org/10.1126/sciadv.adl6078}{Science Advances \textbf{10}, eadl6078 [2024]} under the terms of the \href{https://creativecommons.org/licenses/by/4.0/}{Creative Commons Attribution License 4.0 [CC BY]}.}

\begin{abstract}
What time does a clock tell after quantum tunneling?
Predictions and indirect measurements range from superluminal or instantaneous tunneling to finite durations, depending on the specific experiment and the precise definition of the elapsed time.
Proposals and implementations utilize the atomic motion to define this delay, even though the inherent quantum nature of atoms implies a delocalization and is in sharp contrast to classical trajectories. 
Here, we rely on an operational approach: we prepare atoms in a coherent superposition of internal states and study the time read off via a Ramsey sequence after the tunneling process without the notion of classical trajectories or velocities.
Our operational framework (a) unifies definitions of tunneling delay within one approach; (b) connects the time to a frequency standard given by a conventional atomic clock which can be boosted by differential light shifts; and (c) highlights that there exists no superluminal or instantaneous tunneling.

\end{abstract}

\maketitle
\begin{bibunit}[apsrev4-1mod]
\noindent

\Sec{Introduction}
In relativity, proper time is operationally defined as the time~\cite{einstein1905} measured by an ideal clock~\cite{moller1956} traveling along a specific worldline through spacetime.
However, quantum mechanics in principle allows for motion in classically forbidden regions, culminating in quantum tunneling~\cite{ankerhold2007}.
Hence, the identification of a worldline is intricate, to say the least. 
Although a region may be forbidden for classical motion, the arrival of particles on the other side of a barrier is still observed.  
One can associate the so-called \emph{arrival time}~\cite{buettiker1982} with the appearance of a tunneled particle, giving rise to observations of superluminal~\cite{enders1992, steinberg1993, spielmann1994} and even instantaneous tunnel times~\cite{pfeiffer2013, eckle2008}. 
Since it is impossible to assign classical wordlines to tunneled and delocalized quantum particles, identifying the elapsed time seems to lie outside of the scope of a naive combination of quantum mechanics and general relativity.
Similar to general relativity, we thus follow an operational approach:
reading-off the tunneling time directly from a quantum particle with internal structure that has tunneled through an optical barrier via a Ramsey sequence.

When assigning a time to the arrival of particles on the other side of a barrier, one has to develop concepts due to their quantum nature associated with the Heisenberg uncertainty principle, even without the intricacies of combining general relativity and quantum tunneling.
In fact, particles have to be described by wave packets that are inherently delocalized.
One strategy is to compare the center-of-mass positions of a tunneled particle and a free particle with the same kinetic energy.
Together with the group velocity that can be extracted for a wave packet, the so-called Wigner phase time~\cite{wigner1955, buettiker1983, hauge1989}, or closely related, the group delay can be inferred.
Such an approach is the most common way to define the arrival time, which diverges for small probabilities of tunneling~\cite{buettiker1983}, so that superluminal times have been observed experimentally, \eg, in Hong-Ou-Mandel-type experiments~\cite{steinberg1993, spielmann1994}.

Another technique to measure arrival times is employed in strong-field ionization~\cite{zimmermann2016}.
Attoclock experiments~\cite{uiberacker2007,sainadh2020, pfeiffer2013, eckle2008} assign a time to the electron escape from a bound state of an atom, induced by elliptically polarized light and by that time-varying barrier potentials.
In such experiments, the tunneling time can be inferred from the scattering angle, leading to claims of instantaneous tunneling~\cite{pfeiffer2013, eckle2008}. 
In such experiments, the determination of a group delay depends on the underlying ionization model that incorporates the complex structure of the atom and associated effects~\cite{sainadh2019}.
The inferred arrival time depends therefore strongly on the underlying theoretical modeling~\cite{sainadh2020}.

In contrast to assigning a time to the arrival of particles, the \emph{interaction time} aims at measuring the time a particle spends inside a forbidden region.
This complementary approach gives analytical expressions of a dwell time~\cite{smith1960, buettiker1983, hauge1989} that take into account the average number of atoms inside the barrier and the incoming particle flux.  
While the Wigner phase time diverges for low tunneling probabilities, the dwell time takes a finite value~\cite{buettiker1983, winful2003}. 
One strategy to measure the dwell time is sending a particle onto the barrier and triggering or initiating a clock inside the forbidden region.
Here, the particle is in an internal superposition while tunneling, whereas there is no clock outside the barrier.
The most prominent example is the Larmor clock~\cite{buettiker1983, ramos2020, spierings2021, suzuki2023}.
Here, the degeneracy of two spin states is lifted by a magnetic field that spatially overlaps with the barrier.
A time spent in the forbidden region can be read off by the angle of spin precession.
In this case, the frequency of the ``clock'' is given by the Larmor frequency and depends on the magnetic field and is not connected to a frequency standard.
While instructive for rectangular barriers, it is not clear how other barrier profiles, \eg{}, Gaussian barriers, have to be treated or truncated.

\begin{figure*}
	\centering
	\includegraphics[width=1.0\textwidth]{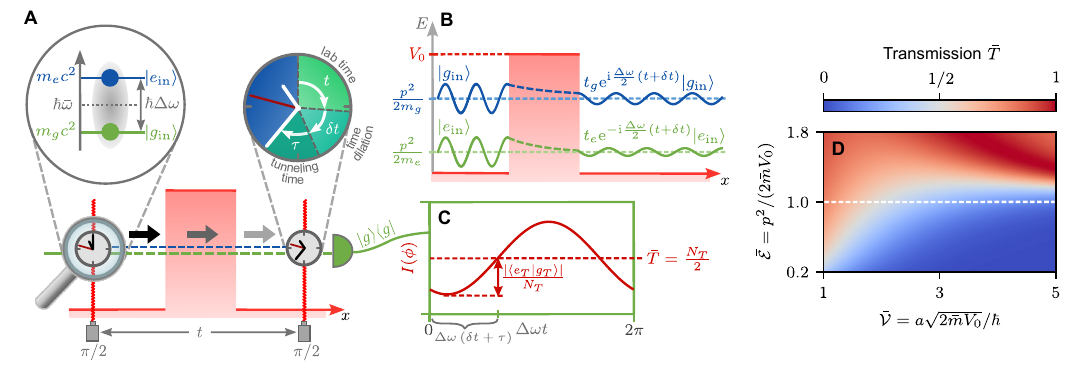}
	\caption{
	\textbf{Tunneling progress of a Ramsey clock}. 
	\textbf{(A)} The first $\pi/2$ pulse of a Ramsey sequence initializes the clock by creating an equal superposition of the internal states $\ket{g_{\text{in}}/e_{\text{in}}}$ of a two-level system.
	Both states are associated with different rest masses $m_{g/e} c^2$ and the energy structure is given by the  clock frequency $\Delta \omega$ (see the magnification). 
	\textbf{(B)} During tunneling, each internal state acquires a state-dependent phase shift encoded in the complex transmission amplitudes $t_{g/e}$.
	After the scattering process, a second $\pi/2$ pulse reads out the accumulated phase which includes contributions from the lab time $t$, time dilation $\delta t$ and tunneling time $\tau$ (see \textbf{(A)}).
	For different lab times, the population in the ground state is detected and an interference signal between both internal states. \textbf{(C)} is obtained, characterized by the contrast $|\braket{e_T}{g_T}| / N_T$ with the total number of transmitted atoms $N_T$ and mean transmission coefficient $\bar{T} = N_T / 2$.
	\textbf{(D)} For a rectangular barrier, this transmission coefficient shows distinct features for different scaled kinetic energies $\bar{\mathcal{E}}$ and dimensionless barrier parameters $\bar{\mathcal{V}}$.
	} 
	\label{fig:fancy_clock_illustration}
\end{figure*}

In contrast to the previous approaches and in the operational spirit of general relativity, we propose measuring the interaction time with a Ramsey clock~\cite{nicholson2015, brewer2019, oelker2019}, consisting of an atom with internal structure that possesses clock degrees of freedom also in the asymptotic region far away from the barrier.
In this case, no external trigger is necessary to start the time measurement within the barrier, since the time delay and imprinted phase is solely caused by the forbidden region itself.
In following, we refer to a quantum particle with internal structure, \ie{} an atom, where the time is read off via a Ramsey sequence. 
After reflection from the barrier, the Ramsey clock consists of a superposition of internal and external degree of freedoms.
In particular, our proposed Ramsey clock represents the reference of a conventional atomic clock which generally also contains an oscillator, a method for readout, and a counter.
In the following, when we refer to a Ramsey clock, we keep in mind the additional elements needed to obtain an atomic clock.
Moreover, the tunneling process itself is probabilistic and performing the proposed experiments on a single-atom level will thus be dominated by quantum projection noise.
We therefore understand all results imply either a repetition of a single-atom experiment, or a preparation of a cold cloud of a large number of identical atoms, maybe even a Bose-Einstein condensate.
Of course, the latter implementation is more feasible due to the statistics, as we discuss below.

Our idealized Ramsey clock consists of two internal states $\ket{e/g}$ with respective eigenenergies $\hbar \omega_{e/g}$ and a frequency standard given by the clock frequency $\Delta \omega = \omega_{e} - \omega_{g}$ that corresponds to the transition frequency, see the magnification in Fig.~\ref{fig:fancy_clock_illustration}\hyperref[fig:fancy_clock_illustration]{A}.
Due to the relativistic mass defect~\cite{yudin2018,sonnleitner2018,schwartz2019,martinez2022,assmann2023}, the mass of an atom in internal state $\ket{e/g}$ 
\begin{equation}
\label{eq:mass_defect}
    m_{e/g} = \bar{m} \left( 1 \pm \frac{\Delta \omega}{2\bar{ \omega}} \right)
\end{equation}
depends explicitly on the state, leading to relativistic effects of the motion and a coupling of external and internal degrees of freedom.
Here, the mean frequency $\bar \omega = (\omega_e+\omega_g)/2 = \bar m c^2/\hbar $ can be connected to the mean mass that dictates the atomic motion to lowest order.

Both internal and external dynamics of such a Ramsey clock without transitions or loss channels induced by the optical potential are described~\cite{yudin2018,sonnleitner2018,schwartz2019,martinez2022,assmann2023} by the two-level Hamiltonian $\hat{H} = \sum_{j=g,e} \hat{H}_j \ket{j}\bra{j}$, where
\begin{align}
    \hat{H}_{e/g} = m_{e/g} c^2 + \frac{\hat{p}^2}{2 m_{e/g}} + V_{e/g}(\hat{x}) 
    \label{eq:int_hamiltonian}
\end{align}
describes the motion of the atom in state \ket{e/g}.
Here, the position and momentum operators $\hat{x}$ and $\hat{p}$ fulfill the commutation relation $\left[\hat{x}, \hat{p} \right] = \ii \hbar$.
The Hamiltonian from equation~\eqref{eq:int_hamiltonian} consists of three parts:
(i) The rest energy giving rise to the clock phase accumulated by each internal state upon time evolution.
(ii) The mass defect included in the kinetic energy, which implies state-dependent dispersion relations and introduces time-dilation effects.
(iii) A possibly state-depended barrier $V_{e/g}(\hat{x})$ , \eg{}, induced by Stark shifts from far-detuned optical fields, including no gravitational contributions. 
In the following, we utilize this relativistic extension of standard quantum mechanics to describe the motion and quantum tunneling of a Ramsey clock and show how to infer the tunneling time.

\Sec{Results}
To measure the phase difference between both internal states after tunneling, a Ramsey sequence~\cite{ramsey1950} is \emph{the} method of choice and depicted in Fig.~\ref{fig:fancy_clock_illustration}\hyperref[fig:fancy_clock_illustration]{A}.
In such a sequence, the Ramsey clock is initialized by a $\pi/2$ pulse at time $t=0$ that generates an equal superposition of both internal states, before the clock impinges on a short-range potential and partially tunnels to the other side.
The phase difference measured by the tunneled clock is read out by a second $\pi/2$ pulse after some time $t$, mixing again both internal states.
The population in the ground state after the Ramsey sequence for perfect instantaneous pulses give rise to a interference signal that is shown in Fig.~\ref{fig:fancy_clock_illustration}\hyperref[fig:fancy_clock_illustration]{C} and takes the form
\begin{equation}
    I = \frac{N_T}{2} \left[ 1 + \frac{|\braket{e_T}{g_T}|}{N_T} \operatorname{cos}\left( \operatorname{arg} \braket{e_T}{g_T} \right)\right]
\end{equation}
where $\ket{g_T}$ and $\ket{e_T}$ are the states of the atom transmitted in the ground and excited state, respectively. 
The amplitude of the interference signal is the mean number of transmitted atoms $N_T = (\braket{e_T}{e_T} + {\braket{g_T}{g_T}}) / 2$ and the contrast is described by the overlap $|\braket{e_T}{g_T}| / N_T$.

The phase difference measured by the tunneled clock
\begin{equation}
    \operatorname{arg} \braket{e_T}{g_T} = 
    \Delta \omega (t + \delta t + \tau) - \left[ \phi(t) - \phi(0) \right]
    \label{eq:phase_diff}
\end{equation}
contains the \emph{laboratory time $t$} given by the separation time of Ramsey fields.
It is measured by the clock frequency $\Delta \omega$.
\emph{Relativistic time dilation}~\cite{einstein1905, pikovski2017, paige2020} induced by the state-dependent dispersion relation results in a modification of the laboratory time by $\delta t$. 
However, our focus lies on the third contribution caused by tunneling through the potential barrier itself: the \emph{tunneling time} $\tau$. 
Additional phase contributions $\phi(t)$ arise from the laser pulses, that can be used to lock the Ramsey fields to the transition frequency and read out the signal.

Assuming that both internal states $\ket{g_\text{in}}$ and $\ket{e_\text{in}}$ have the same initial momentum, the transmitted states are depicted in Fig.~\ref{fig:fancy_clock_illustration}\hyperref[fig:fancy_clock_illustration]{B} and are given by
\begin{equation}
    \ket{e_T/g_T} = t_{e/g} \exp{\left[ -\ii \left( \omega_{e/g} t - \frac{p^2}{2 m_{e/g} \hbar} t \right) \right] } \ket{e_\text{in}/g_\text{in}},
\end{equation}
see methods for details.  
The transmission amplitudes $t_{e/g}$ contain all information on the tunneling process and are complex quantities with $|t_{e/g}|\leq 1$.
In addition to the phases $\operatorname{arg}t_{e/g}$ induced by tunneling, other contributions are associated with the internal energy and the kinetic energy of the atom.

The phase imprinted by kinetic energy depends explicitly on the mass of the atom and, by means of the mass defect introduced in equation~\eqref{eq:mass_defect}, implies time dilation.
To lowest order, this contribution leads to a time delay
\begin{equation}
    \delta t = \frac{1}{2} \left(\frac{p}{\bar{m} c} \right)^2 t .
    \label{eq:time_dilation_eig}
\end{equation}
It arises independently of tunneling and is measured by any moving Ramsey clock~\cite{paige2020}.

\begin{figure}[ht]
	\begin{center}
		\includegraphics[width=0.5\textwidth]{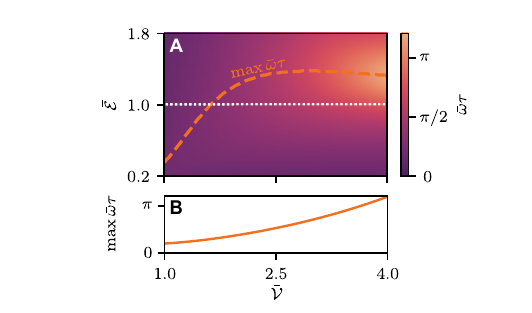}
		\caption{
		\textbf{Tunneling time for a rectangular barrier.}
		\textbf{(A)}~Scaled tunneling time $\bar{\omega} \tau$ for different kinetic energies $\bar{\mathcal{E}}$ and barrier parameters $\bar{\mathcal{V}}$.   
		For opaque barriers, the maximal tunneling time (orange dashed line) moves from the tunneling regime ($\bar{\mathcal{E}} < 1$) to the classical regime ($\bar{\mathcal{E}} > 1$) and then approaches the barrier height ($\bar{\mathcal{E}} = 1$).
		\textbf{(B)}~The maximal tunneling time increases for an increasing barrier parameter.   
		} 
		\label{fig:tun_phase_rec}
	\end{center}
\end{figure}

Before turning to more general barriers, we study the phase measured by a clock that tunnels through a rectangular barrier with height $V_0$ and length $a$.
To highlight the relevant quantities, we introduce a dimensionless kinetic energy $\bar{\mathcal{E}} = \bar{E} / V_0 = p^2 / (2 \bar{m} V_0)$ that is smaller than unity if the clock is tunneling.
Moreover, we define the dimensionless barrier parameter as $\bar{\mathcal{V}} = \int \dd x \sqrt{2 \bar{m} V(x)} / \hbar$. 
For $\bar{\mathcal{V}} \gg 1$ we have an opaque barrier, while for $\bar{\mathcal{V}} \ll 1$ the barrier is transparent. 
In the case of rectangular barriers the expression simplifies to $\bar{\mathcal{V}} = a \sqrt{2 \bar{m} V_0} / \hbar$.
We observe that the product $\bar m V_0$ always enters both parameters.
Thus, state-dependent barrier heights or the relativistic mass defect will induce similar effects on tunneling.
In fact, this feature connects our work to the concept of Larmor clocks~\cite{buettiker1983} and already shows that the effect of a mass defect can be mimicked by state-dependent barriers.

By expanding the transmission amplitudes of ground and excited states to first order in $\Delta \omega / \bar \omega$, we find the tunneling time
\begin{equation}
    \tau =
    \frac{ \bar{\mathcal{V}} \bar{T}}{4 \bar{\omega} \sqrt{\bar{\mathcal{E}}} (\bar{\mathcal{E}} - 1)}
    \left[\left( 2 \bar{\mathcal{E}} - 1 \right) - \frac{\operatorname{sinh}\left( 2 \bar{\mathcal{V}} \sqrt{1 - \bar{\mathcal{E}}} \right)}{ 2 \bar{\mathcal{V}} \sqrt{1 - \bar{\mathcal{E}}}}\right].
    \label{eq:tun_time_eig}
\end{equation}
It is proportional to the mean transmission coefficient 
\begin{equation}
    \bar{T} = \left[ 1 - \frac{1}{4 \bar{\mathcal{E}} (\bar{\mathcal{E}} - 1)} \operatorname{sinh}^2 \left(\bar{\mathcal{V}} \sqrt{1 - \bar{\mathcal{E}}}\right) \right]^{-1}
\end{equation}
shown in Fig.~\ref{fig:fancy_clock_illustration}\hyperref[fig:fancy_clock_illustration]{D} for different $\bar{\mathcal{E}}$ and $\bar{\mathcal{V}}$, which causes shorter tunneling times far below the barrier where tunneling processes are less likely. 
In contrast, for $\bar{\mathcal{E}}\gg 1$, that is for traveling clocks that hardly feel the barrier and do not tunnel, the delay $\tau$ vanishes.  
Connecting both asymptotics shows that there exists a maximal tunneling time, as highlighted in Fig.~\ref{fig:tun_phase_rec}\hyperref[fig:tun_phase_rec]{A} by the dashed line in parameter space.
For transparent barriers $ \bar{\mathcal{V}}< 1.5$ the maximum is achieved for $\bar{\mathcal{E}} < 1$, while for opaque barriers $ \bar{\mathcal{V}}\gg 1$ the maximal delay is achieved for $\bar{\mathcal{E}} = 1$.
Nevertheless, the density plot shows a smooth transition from $\bar{\mathcal{E}} < 1$ to $\bar{\mathcal{E}} > 1$. 
Thus, there is no clear distinction between a tunneling clock and one that travels classically above the barrier. 
Moreover, Fig.~\ref{fig:tun_phase_rec}\hyperref[fig:tun_phase_rec]{A} shows that the tunneling time increases for opaque barriers.

\begin{figure}[ht]
	\begin{center}
		\includegraphics[width=0.5\textwidth]{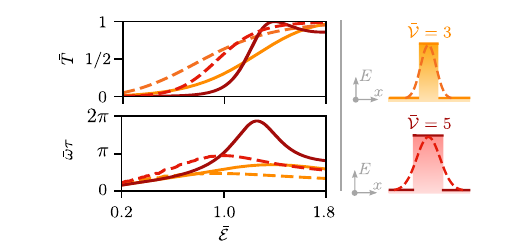}
		\caption{
		\textbf{Transmission and tunneling time for different types of barriers.}
		Comparison of the mean transmission coefficient $\bar{T}$ and scaled tunneling time $\bar{\omega} \tau$ of two rectangular (solid line) and Gaussian barriers (dashed line) for different scaled kinetic energies $\bar{\mathcal{E}}$.
		The transmission amplitude for plane waves scattered on Gaussian barriers is obtained by the transfer matrix approach.
        For the calculations, we have assumed Gaussian barriers of height $k_B \times 200 \, \mathrm{nK}$~\cite{spierings2021} and $^{87}\mathrm{Rb}$ atoms with mass $m =  86.91 \, u$.
		For Gaussian barriers, the maximal tunneling time is smaller than for rectangular barriers and is shifted to smaller initial kinetic energies. 
		} 
		\label{fig:tun_phase_comp_rec_gauss}
	\end{center}
\end{figure}

To connect to a possible experimental implementation, we extend our study to Gaussian barriers, experimentally achievable, \eg{}, by spatial light modulators, painted potentials or digital micromirror devices~\cite{amico2021}. 
To ensure a proper comparison between such barriers and rectangular potentials, we always choose the same barrier parameter $\bar{\mathcal{V}}$. 
For Gaussian barriers, we fix its height and vary the width to adjust the barrier parameter, as shown in the pictograms on top of Fig.~\ref{fig:tun_phase_comp_rec_gauss}. 
The figure compares the transmission probability and the tunneling time for different rectangular and Gaussian barriers, where the transmission amplitudes are determined semi-analytically by a transfer matrix ansatz~\cite{jirauschek2009, loran2020}. 
For $\bar{\mathcal{E}} > 1$ we observe resonances in the transmission probability resulting from partial reflection and transmission of the initial state. 
This structure washes out for Gaussian barriers and no prominent resonances in transmission are observed. 
Of course this behavior is also reflected in the phase obtained during tunneling, and by that also in the tunneling time.
In fact, the tunneling delay induced by Gaussian barriers is smaller than for rectangular barriers with the same barrier parameter.
Moreover, the maximal tunneling time moves to lower initial energies.

So far we considered tunneling times associated to specific momentum eigenstates.
However, since Ramsey clocks are localized quantum objects, the Heisenberg uncertainty relation implies that we in fact have to consider a momentum distribution, \eg, in simple cases a (Gaussian) wave packet  $\psi_0(p) = \braket{p}{e/g(0)}$.
As a consequence, the tunneling time becomes explicitly momentum dependent, \ie, we replace $\tau \to \tau(p)$, because the kinetic energy $\mathcal{\bar{E}} \to \mathcal{\bar{E}}(p)$ and by that the transmission probability $\bar{T} \to \bar{T}(p)$ both depend on momentum.
Considering the mass defect to first order, we still observe perfect contrast $|\braket{e_T}{g_T}|/N_T=1$ of the Ramsey fringe.
However, the induced time dilation for of a tunneled wave packet takes the form
\begin{equation}
    \delta t = \frac{t}{2 N_T} \int \dd p \bar{T}(p)|\psi_0(p)|^2 \left( \frac{p}{\bar{m} c} \right)^2.
    \label{eq:avg_time_dilation}
\end{equation}
Following the same line of argument, the tunneling time, defined in equation~\eqref{eq:tun_time_eig} for a momentum eigenstate, is generalized to
\begin{equation}
    \tau = \frac{1}{N_T} \int \dd p \bar{T}(p)|\psi_0(p)|^2 \tau(p).
    \label{eq:avg_time_tunneling}
\end{equation}
Both times describe a momentum average of the eigenstate solution over the tunneled momentum distribution. 
Consequently, the average tunneling time is smaller for wave packets than for momentum eigenstates.
Hence, to measure a large time delay in an experimental implementation, collimated wave packets with ultralow expansion rates and momenta that correspond to energies slightly above the barrier height are desirable.
Techniques like delta-kick collimation that achieve momentum widths in the order of picokelvin~\cite{ammann1997}, in combination with magic~\cite{katori2003} Bragg diffraction~\cite{giese2015} might be one option to observe these tunneling times experimentally.

\Sec{Discussion}

\begin{figure}[ht]
	\begin{center}
		\includegraphics[width=0.5\textwidth]{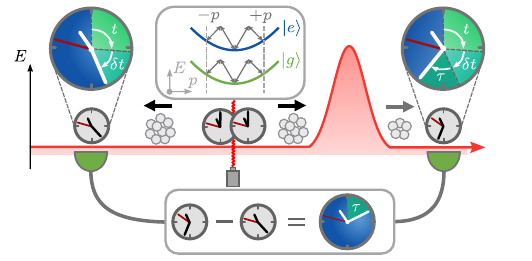}
		\caption{
		\textbf{Differential measurement-scheme for isolating the tunneling phase.}
		To start the protocol, we prepare an atomic cloud and initialize a Ramsey clock by generating a superposition of two internal states $\ket{g/e}$.
		Subsequently, we create an equal superposition of two Ramsey clocks with opposite momenta $\pm p$, realized by (magic) double Bragg diffraction as shown in the inset.
        While one clock tunnels, the other one is used as reference to cancel common phase contributions, \eg, the clock phase $\Delta \omega t$ and the time-dilation phase $\Delta \omega \delta t$.
        From the remaining phase, one can deduce the tunneling time $\tau$.
		} 
		\label{fig:diff_mesurement}
	\end{center}
\end{figure}

To isolate the phase accumulated during tunneling, we propose a differential measurement scheme, visualized in Fig.~\ref{fig:diff_mesurement}.
As a first step, we prepare an atomic cloud with finite momentum width.
Subsequently, the Ramsey clock is initialized by generating an internal superposition of each atom in the cloud, using preferably recoilless transitions like $E1M1$~\cite{alden2014, janson2023} or operating in the Lamb-Dicke regime~\cite{dicke1953} and releasing the clock after initialization.
Subsequently, we create a superposition of of the clock moving with opposite momenta, \eg{}, realized via double Bragg diffraction~\cite{giese2015, giese2013} at the magic wave length~\cite{katori2003} to ensure that both internal states are diffracted equally.
While one clock tunnels, the second one serves as a reference to cancel common phase contributions,\eg{}, time dilation, the laboratory time, and phase noise of interrogating Ramsey fields. 
As a result, the differential phase measurement between the tunneled and reference clock provides the tunneling time. 

Since we measure the ground-state populations of the tunneled and reference Ramsey clocks, position uncertainties of the atomic cloud and thus different arrival times play no role for detection.   
Nevertheless, velocity uncertainties due to the finite momentum width $\delta p$ of the wave packet has to be considered and in principle leads to contributions in equations~\eqref{eq:avg_time_dilation} and \eqref{eq:avg_time_tunneling}.
However, for an atomic cloud centered around a momentum $p_0$, a Taylor expansion shows that for $|\partial_p \bar T|_{p_0} \delta p| \ll 1$ the transmission coefficient $\bar T$ of the integrand cancels due to the normalization $N_T$.
In this case, both the tunneled Ramsey clock and the reference clock observe the same time dilation $\delta t$, so that only the phase contribution caused by the finite tunneling time survives in a differential measurement.
Figure~\ref{fig:fancy_clock_illustration}D allows us to identify a parameter regime $\bar{\mathcal{E}} > 1$ where $\bar{T} \approx 1$ is maximal and by that approximately constant, so that the condition $| \partial_p \bar T|_{p_0} \delta p| \ll 1$ is satisfied for feasible momentum distributions.
Luckily, according to Fig.~\ref{fig:tun_phase_rec} this is the parameter regime where for opaque barriers $\bar{\mathcal{V}} \gtrsim 2$  the tunneling time is maximal, and the experiment is optimally operated.
Since techniques like delta-kick collimation~\cite{kovachy2015, muntinga2013, gaaloul2022} have achieved momentum uncertainties that correspond to the picokelvin temperatures, time dilation contributions can be suppressed by such a differential measurement.

However, while such a setup may isolate the tunneling time, additional phases may arise in an actual experimental realization, see the method section for analytical expressions of some of the spurious effects. 
For example, imperfect preparation of the center-of-mass motion of the Ramsey clock, like an initialization without recoilless transitions or outside the Lamb-Dicke regime, may lead to state-dependent initial momenta.
If both clock states have different polarizabilites or couple differently to the optical potential, they may experience state-dependent barrier heights.
Both effects are discussed in the following.

Phase contributions arising from state-dependent initial momenta take the form of a differential Doppler shift.
Key to a suppression is, as mentioned above, that the excited state has exactly the same momentum as the ground state.
When initializing the clock, the photon recoil imparted onto the atom upon absorption directly introduces such an effect.
For microwave transitions, this effect is negligible due to the dispersion relation, however, the clock frequency is also decreased by this factor.
In contrast, the photon recoil of optical clock transitions is nonnegligible and actually used for atom interferometry~\cite{kasevich1991, cronin2009}.
One way to suppress the recoil imparted during initialization of the clock, the experiment has to be performed in a sufficiently confining trap within the Lamb-Dicke regime~\cite{dicke1953}.
Following such a strategy, one still has to analyze the trap release to avoid imprinting additional clock phases or accelerating the atom.
As a second step, one has to impart the same momentum to both states.
One possibility are Bragg gratings which have the benefit that the transferred momentum depends on the effective wave vector of the grating and is independent of the state.
One only must ensure that the effective Rabi frequency is the same for both internal states, \eg{}, by operating at the magic wavelength.
Another approach is to apply Doppler-insensitive and recoilless two-photon transitions to an atom at rest or moving in the ground state, such as $E1M1$ transitions, so that the release from the trap does not pose an issue.
However, the differential Doppler effect between both internal states gives additional insight into the tunneling process.
We show in the method section that it is directly connected to a measurement of the arrival time~\cite{buettiker1982} on the scale of the differential Doppler frequency.
In contrast to existing models that track the peak of a wave packet, \eg{} through Hong-Ou-Mandel interference~\cite{steinberg1993, spielmann1994}, our proposed scheme does not rely on a measurement of the peak itself, since the group delay is encoded in the Ramsey fringes.
It actually does neither rely on the notion of peaks nor on assigning a classical velocity to wave packets or a clock.  

In addition to imperfect preparation of the initial momenta, another spurious phase contribution may arise from state-dependent barriers.
This phase is inherently connected to the concept of Larmor clocks~\cite{buettiker1983, ramos2020, spierings2021, suzuki2023}, which describe the tunneling of a superposition of spin states with degenerate eigenenergies.
This degeneracy is lifted by a magnetic field only present in the barrier region, which initiates or triggers the Larmor clock.
In an alternative description, the different eigenenergies are equivalent to spin-dependent barrier heights, \ie{}, the process is equivalent to a perturbation for our proposed experiment. 
While the degeneracy is lifted by the magnetic field, that is, during the tunneling process, spin precession induces a phase.
In this case, the Larmor frequency at which the interaction time is probed, can be tuned by the magnetic field.  
In contrast to our treatment, there are no relativistic corrections included in such a treatment.

So far, only experiments~\cite{ramos2020, spierings2021} realizing the analogue of Larmor clocks have been implemented. 
In these cases, the two spin states are replaced by two internal atomic states and the magnetic field by an external electromagnetic field that induces Raman transitions inside the barrier region, so that the Rabi frequency plays the role of the Larmor frequency and can be tuned by the intensity of the Raman beams.
As a consequence, the analogue of spin precession is the effective Rabi oscillation between both internal states.
However, no superposition of internal states is prepared initially, so that the implementation of such a Larmor clock does not correspond to a Ramsey clock scheme in a strict sense. 
Moreover, a Ramsey clock ticks at the intrinsic frequency given by the energy difference of both states, whereas the Larmor-clock analogue ticks typically at Rabi frequencies in the microwave regime and thus does not correspond to a good frequency standard. 

As described above, the interaction time $\tau_L$ can be measured by spin-dependent barriers.
We therefore compare the magnitude of the phase contribution measured in Larmor-clock experiments to the ones that arise from the mass defect and find the ratio
\begin{equation}
   \frac{\tau_L \omega_L}{\tau \Delta \omega} = \frac{\hbar \bar{\omega}}{\bar{V}} \frac{\omega_L}{\Delta \omega} \approx 3.1 \times 10^{14},
\end{equation}
where the expansion coefficients in both cases are the same and cancel, as discussed in the methods section.
Based on the experimental implementation~\cite{spierings2021}, we use the clock frequency of $\Delta \omega = 2 \pi \times 6.8\, \mathrm{GHz}$ related to the $F=1$ to $F=2$ transition of the state $5\mathrm{S}_{1/2}$ of $^{87}\mathrm{Rb}$ with mass $m =  86.91 \, u$ and the atomic mass unit $u$.
The mean barrier height is $\bar{V} = 1.3 \times 10^{-31}\, \mathrm{J}$~\cite{spierings2021} and the differential barrier height corresponds to the Rabi frequency $\Delta V/\hbar =  \omega_L = 2 \pi \times 200\, \mathrm{Hz}$~\cite{spierings2021}.
In the experiments performed, the main phase contribution comes from the Larmor clock that is artificially imprinted  and triggered by the state-dependent barrier.

On the other hand, the clock frequency between two hyperfine states of rubidium is in the gigahertz regime and by no means an optical frequency that enhances the measured phase.
Ideally one should perform the proposed experiment on such an optical clock transition.
In our setup, imperfections and perturbations induced by state-dependent barriers lead, in analogy to Larmor clocks, to an additional phase that depends on the tunneling time.   
In fact, potential barriers where each state of the clock transition experiences a different light shift and by that potential, will induce such a phase.
Since both contributions have the same expansion coefficient, an effective clock frequency can be identified  
\begin{equation}
     \omega_{\mathrm{eff}} = \Delta \omega \left( 1 + \frac{\bar{\omega}}{\Delta \omega} \frac{\Delta V}{\bar{V}} \right) .
     \label{eq:eff_clock_freq}
\end{equation}
To suppress the Larmor phase and isolate the relativistic clock contribution, one has to generate the optical barrier from light at the magic wavelength, so that we have $(\Delta V /\bar{V}) \ll (\Delta \omega / \bar{\omega})\cong 1.33 \times 10^{-11}$.
Here, we have assumed a cold cloud of $^{174}\mathrm{Yb}$~\cite{bouganne2017} atoms with the clock transition $\Delta \omega = 2 \pi \times 522 \, \mathrm{THz}$ and Compton frequency $\bar{\omega} = 2\pi \times 3.92 \times 10^{25} \, \mathrm{Hz}$.
An experimental realization of the differential measurement would be performed in regions of large tunneling times and high transmissions which is achieved at $\bar{\mathcal{V}} \approx 4$ and $\bar{\mathcal{E}} \approx 1.4$. 
Consulting figures~\ref{fig:fancy_clock_illustration} and \ref{fig:tun_phase_rec}, we obtain $\bar{T} \approx 1$ and $\bar{\omega} \tau \approx \pi$ at the optimal working point, resulting in the miniature tunneling time $\tau \approx 1.3 \times 10^{-26}\, \mathrm{s}$, or tunneling phase $\tau \Delta \omega \approx 4.2 \times 10^{-11}$, respectively.
Obviously, such timescales are to date inaccessible to high-precision clocks.
For instance, if we assume a shot-noise limited sensitivity and $10^{5}$ atoms in the ytterbium cloud, one would need repetitions of order $10^{15}$ to resolve the tunneling phase, beyond any feasible implementation.
Even though relying on squeezing~\cite{robinson2024} constitutes one way to reduce the number of runs it still is enormous even for ambitious assumptions.
However, the tunneling phase can be increased considerably by boosting the clock frequency artificially.
As shown in equation~\eqref{eq:eff_clock_freq}, the effective clock frequency depends on the differential barrier heights which can be experimentally realized by differential light shifts.
The mean barrier barrier height $\bar{V} \approx 3.8 \times 10^{-30}\, \mathrm{J}$ is fixed through our choice of the working point and the effective wave number $k = 1.7 \times 10^{7} \, \mathrm{1/m}$ of the magic two-photon double Bragg transition at the magic wavelength $759.35 \, \mathrm{nm}$ \cite{barber2008}.
Differential light shifts in the order of hertz~\cite{barber2008} will already boost the clock frequency by factor $2 \times 10^{6}$, resulting in a tunneling phase of $\tau \Delta \omega_\text{eff} \approx  10^{-4}$.
Assuming a shot-noise limited sensitivity, we obtain a feasible number of runs, \ie{}, of order $10^{3}$ suitable to resolve the miniature tunneling time.

In conclusion, based on an operational approach and in analogy to general relativity, we demonstrated that a tunneled Ramsey clock acquires a phase shift and tells a time that can be associated with the tunneling process.
This time is read out by a preferably optical frequency standard that is independent of any fields triggering the measurement inside the barrier and originates from relativistic effects and properties of the atoms.
Moreover, the clock frequency can be artificially increased by differential light shifts to resolve the tunneling time.
Other approaches to identify a dwell time and an arrival time are also contained in our proposed experiment and can be enhanced by introducing perturbations. 

\Sec{Materials and Methods}
Neglecting the rest-energy term from equation~\eqref{eq:int_hamiltonian}, the time-independent Schr\"odinger equation for two decoupled internal states is given by
\begin{equation}
    \left[ - \frac{\hbar^2}{2 m_j} \frac{\partial^2}{\partial x^2} + V_j(x) \right] \ket{j} = E_j \ket{j}, 
\end{equation}
where $E_j = p_j^2 / (2 m_j)$ is the eigenenergy of particle $\ket{j}$ that can be associated with an eigenmomentum $p_j$ and mass $m_j$.
Here,
\begin{equation}
    V_j(x) = 
    \left\{ \begin{array}{ll}
        V_j & 0 \leq x \leq a \\
        0 & \text{otherwise}
    \end{array} \right.
    \label{eq:rec_pot}
\end{equation}
is a, possibly state-dependent, rectangular barrier.
To obtain the transmission amplitude associated with tunneling through the barrier, we solve the Schr\"odinger equation in the three regions defined by equation~\eqref{eq:rec_pot} for each internal state independently and apply the boundary and continuity conditions to connect the individual solutions. 
As a result, for tunneling with energies $E_j < V_j$, we obtain the transmission amplitude
\begin{equation}
    \label{eq:t_rec}
    t_j = \frac{4 \ii \kappa_{0, j} \kappa_{1, j} \expUp{\ii a \kappa_{0, j}}}
    {\left( \kappa_{0, j} + \ii \kappa_{1, j} \right)^2 \expUp{a \kappa_{1, j}}-
    \left( \kappa_{0, j} - \ii \kappa_{1, j} \right)^2 \expUp{-a \kappa_{1, j}}},
\end{equation}
where we introduced the wave numbers $\kappa_{0, j} = \sqrt{2 m_j E_j} / \hbar$ and $\kappa_{1, j} = \sqrt{2 m_j (V_j - E_j)} / \hbar$ associated with the initial kinetic energy and the tunneling process.
To find the transmission amplitude for $E_j > V_j$, we replace $\kappa_{1, j} \to \ii \sqrt{2 m_j (E_j - V_j)} / \hbar$ in the equation above.

\begin{figure}[ht]
	\begin{center}
		\includegraphics[width=0.5\textwidth]{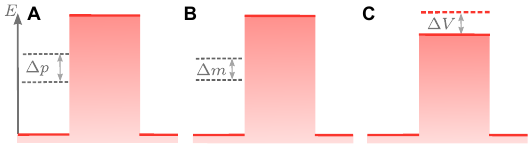}
		\caption{
		\textbf{Perturbations contributing to the interference signal.} 
		These phase contributions may occur from imperfect preparation of the initial momenta $\Delta p$ (\textbf{A}), the relativistic mass defect $\Delta m$ (\textbf{B}) or state-dependent barrier heights $\Delta V$(\textbf{C}).  
		While the mass defect is an intrinsic property, state-dependent momenta and barriers can be mitigated with proper schemes as discussed in the main body of the article. 
		} 
		\label{fig:vis_imperfections}
	\end{center}
\end{figure}
 
For convenience, we introduce the dimensionless kinetic energy $\mathcal{E}_j = E_j / V_j$ and the barrier parameter $\mathcal{V}_j = a \sqrt{2 m_j V_j} / \hbar$.
To study perturbations induced by the mass defect~\cite{loriani2019}, we introduce state-dependent masses $m_{e/g} = \bar{m} \pm \Delta m / 2$.
Moreover, we take into account imperfections in the preparation of the initial state, resulting in state-dependent momenta $p_{e/g} = \bar{p} \pm \Delta p / 2$, and possibly state-dependent barrier heights $V_{e/g} = \bar{V} \pm \Delta V / 2$. 
Together with the mass defect, possible state-dependent momenta and barrier heights are visualized in Fig.~\ref{fig:vis_imperfections}.
Here, we assume that $\Delta m / \bar m$, $\Delta p / \bar p$, and  $\Delta V / \bar V$ can be treated as perturbations.

To first order in all perturbations, the state-dependent transmission amplitude 
\begin{equation}
    T_{e/g} = |t_{e/g}|^2 = \bar{T} \pm T_m \frac{\Delta m}{2 \bar{m}} \pm T_V \frac{\Delta V}{2 \bar{V}} \pm T_p \frac{\Delta p}{2 \bar{p}} + \dots
\end{equation}
fulfills the relations $T_e T_g \approx \bar{T}$ and $(T_e + T_g)/2 \approx \bar{T}$ due to the intrinsic symmetry, where we defined the mean transmission coefficient 
\begin{equation}
    \bar{T} = \left[ 1 - \frac{1}{4 \bar{\mathcal{E}} (\bar{\mathcal{E}} - 1)} \operatorname{sinh}^2 \left(\bar{\mathcal{V}} \sqrt{1 - \bar{\mathcal{E}}}\right) \right]^{-1} .
\end{equation}
Hence, the visibility of the Ramsey fringe is not affected to lowest order of the perturbations.

In a similar manner, we expand the phase $\varphi_j = \operatorname{arg}(t_j)$ of the transmission amplitude given by
\begin{align}
    \varphi_j  =&  
    \operatorname{arctan}\left[\frac{(2 \mathcal{E}_j - 1)\operatorname{tanh}\left( \mathcal{V}_j \sqrt{1 - \mathcal{E}_j} \right)}{2 \sqrt{\mathcal{E}_j} \sqrt{1 - \mathcal{E}_j}}  \right]-\mathcal{V}_j \sqrt{\mathcal{E}_j} 
\end{align}
and obtain, up to the first order in all perturbative parameters, the expression
\begin{subequations}
\begin{equation}
    \varphi_{e/g} = \bar{\varphi} \pm \Phi_m \frac{\Delta m}{2 \bar{m}} \pm \Phi_V \frac{\Delta V}{2 \bar{V}} \pm \Phi_p \frac{\Delta p}{2 \bar{p}} + \cdots.
\end{equation}
The expansion coefficients associated with the mass defect and different potential heights coincide and take the form 
\begin{align}
\begin{split}
    \Phi_m = \Phi_V =&  
    -\frac{1}{8} \frac{\bar{T}}{\bar{\mathcal{E}} (\bar{\mathcal{E}} - 1)} 
    \Bigg[ 2 \bar{\mathcal{V}} \sqrt{\bar{\mathcal{E}}} \left( 2 \bar{\mathcal{E}} - 1 \right)  \\
    \vphantom{=}& - \sqrt{\frac{\bar{\mathcal{E}}}{1 - \bar{\mathcal{E}}}} \operatorname{sinh}\left( 2 \bar{\mathcal{V}} \sqrt{1 - \bar{\mathcal{E}}} \right) \Bigg] 
    \end{split}
\end{align}
In particular, the contribution is negative for all values of $\bar{\mathcal{E}}$ and $\bar{\mathcal{V}}$, so that $\Phi_m = - |\Phi_m|$.
Imperfections in the initial preparation of momenta lead to the expansion coefficient
\begin{align}
\begin{split}
    \Phi_p =  \frac{-\bar{\mathcal{V}}\bar{T}}{8 (\bar{\mathcal{E}} - 1) \sqrt{\bar{\mathcal{E}}}} 
    \Bigg[& 1 - 4 \bar{\mathcal{E}} - \operatorname{cosh}\left( 2 \bar{\mathcal{V}} \sqrt{\bar{\mathcal{E}} - 1} \right) \\
    &+ \frac{2 \operatorname{sinh}\left( 2 \bar{\mathcal{V}} \sqrt{\bar{\mathcal{E}} - 1} \right)}{\bar{\mathcal{V}} \sqrt{\bar{\mathcal{E}} - 1}} \Bigg].
\end{split}
\end{align}
\end{subequations}
The phase induced by two different state-dependent momenta corresponds to a differential Doppler shift $\bar p\Delta p / \bar m $, which is apparent from the relation 
\begin{equation}
    \Phi_p \frac{\Delta p}{\bar{p}} = \Phi_p \frac{1}{2 \bar{\mathcal{E}}} \frac{\bar{p} \Delta p}{\bar{m}} \frac{1}{\bar{V}}.
\end{equation}

Additional phase contributions arise when moving to a tunneled Ramsey clock described by the two internal states with frequencies $\omega_j = \bar{\omega} \pm \Delta \omega / 2$ and their difference $\Delta \omega$.
Including the respective rest energy and relying on the time-dependent Schr\"odinger equation, the transmitted internal states are given by
\begin{equation}
    \ket{j_T} = \sqrt{T_j(p)} \exp \left[ -\ii \left( \omega_j t - \frac{p^2}{2 m_j \hbar} t - \varphi_j \right)\right] \ket{j_\text{in}}
\end{equation}
and include phases from the internal states, the motion of the atoms, and tunneling. 
After a Ramsey sequence, we measure the phase 
\begin{equation}
    \operatorname{arg} \braket{e_T}{g_T} = \Delta \omega (t + \delta t + \tau) + \tau_L \omega_L + \tau_D \frac{\bar{p} \Delta p}{\hbar \bar{m}}
\end{equation}
where possible state-dependent initial momenta and potentials are taken into account. 
The tunneling time arises for small mass defects 
\begin{equation}
    \tau \Delta \omega = \frac{|\Phi_m|}{\bar{\omega}} \Delta \omega.
\end{equation}
As a consequence of the dispersion relation, we obtain a modification of the laboratory time related to time dilation, \ie ,
\begin{equation}
    \Delta \omega \delta t = \frac{\Delta \omega}{2} \left(\frac{\bar{p}}{\bar{m} c} \right)^2 t.
\end{equation}
Moreover, imperfect preparation of the initial state may lead to state-dependent initial momenta and gives rise to the differential Doppler time
\begin{equation}
    \tau_D \frac{\bar{p} \Delta p}{\hbar \bar{m}} = \left[ \frac{t}{2} + \tau_P \right] \frac{\bar{p} \Delta p}{\hbar \bar{m}}
\end{equation}
with the Wigner phase time $\tau_P = \hbar \partial_{\bar{E}} \bar{\varphi}= \hbar \Phi_p/ (2 \bar{E})$. 
Consequently, the proposed scheme can be used to measure the Wigner phase time~\cite{wigner1955} with a Ramsey clock when preparing internal states with different momenta. 
Similarly introducing a Larmor frequency $\omega_L= \Delta V / \hbar$ connected to the height difference of the state-dependent barriers, barrier imperfections lead to the Larmor time $\tau_L$ of the form
\begin{equation}
    \tau_L \omega_L = \frac{\hbar |\Phi_V|}{\bar{V}} \omega_L.
\end{equation}

\putbib[TunTimes]

\Sec{Acknowledgements}
We are grateful to A. Friedrich for his stimulating input and proofreading of the manuscript. 
The QUANTUS+ and INTENTAS projects are supported by the German Aerospace Center (Deutsches Zentrum für Luft- und Raumfahrt, DLR) with funds provided by the Federal Ministry for Economic Affairs and Climate Action (Bundesministerium für Wirtschaft und Klimaschutz, BMWK) and the Federal Ministry of Economic Affairs and Energy (Bundesministerium für Wirtschaft und Energie, BMWi) due to an enactment of the German Bundestag under grant no. 50WM2250E (QUANTUS+) and 50WM2177 (INTENTAS). EG thanks the German Research Foundation (Deutsche Forschungsgemeinschaft, DFG) for a Mercator Fellowship within CRC 1227 (DQ-mat).

\Sec{Author contributions}
Conceptualization: EG.
Methodology: PS and EG.
Investigation: PS.
Visualization: PS.
Supervision: EG.
Writing—original draft: PS.
Writing—review \& editing: PS and EG.

\Sec{Competing interests}
The authors declare that they have no competing interests.

\Sec{Data and materials availability}
All data needed to evaluate the conclusions in the paper are present in the paper. 

\end{bibunit}

\end{document}